\documentstyle[aps,preprint]{revtex}
\newcommand{\be}{\begin{equation}}
\newcommand{\ee}{\end{equation}}
\newcommand{\bea}{\begin{eqnarray}}
\newcommand{\eea}{\end{eqnarray}}
\newcommand{\f}{\frac}

\def\6{\partial}
\def\tp{{\tilde{\phi}}^{(\alpha )}}
\def\F{{\cal{F}}^{(\alpha )}}
\def\p{{\phi}^{(\alpha )}}
\def\E{E^{(\alpha )}}
\def\a{\alpha}
\def\b{\beta}
\draft
\begin{document}

\title{Gravity and Duality between Coordinates and Matter Fields}
\author{I. V. Vancea\footnote{e-mail: ivancea@ift.unesp.br,vancea@cbpf.br\\
On leave from Babes-Bolyai University of Cluj, Romania} }
\address{ Instituto de F\'{\i}sica Te\'{o}rica,
Universidade Estadual Paulista\\
Rua Pamplona 145, 01405-900 S\~{a}o Paulo - SP\\
and\\
Grupo de F\'{\i}sica Te\'{o}rica, 
Universidade Catolica de Petr\'{o}polis (UCP)\\
Rua Bar\~{a}o do Amazonas, 124, 25.685-000, Petr\'{o}polis-RJ, Brazil}

\maketitle


\begin{abstract}
We use the duality between the local Cartezian coordinates and the solutions
of the Klein-Gordon equation to parametrize locally the spacetime in terms
of wave functions and prepotentials. The components of metric, metric
connection, curvature as well as the Einstein equation are given in this
parametrization. We also discuss the local duality between coordinates and
quantum fields and the metric in this later reparametrization.

\end{abstract}


\vspace{1cm}

As was recently shown in \cite{fm}, the Cartezian coordinates of flat
spacetime can be interpreted as functionals on solutions of the Klein-Gordon
equation and some other functionals called prepotentials \cite{sw,bm,mm}.
This duality between coordinates and matter fields allows, on one hand, to
give a statistical interpretation to the spacetime coordinates in classical
quantum mechanics \cite{fm} (see also \cite{af,bfm}.) On the other hand, it
suggests that the measurements performed in terms of coordinates could be
expressed in terms of fields, too. This idea might be useful, for example,
at Planck scale where the fields represent practically the unique physical
objects that can play the role of measuring devices. However, at Planck scale
the effects of gravity are rather strong. Therefore, it is important to
understand the duality between coordinates and fields in the presence of
gravity (for recent attempts to formulate the gravity and supergravity
in terms of quantum
quantities see \cite{lr,gl,viv,cv,pv}.) The conditions under which this duality
can hold locally on a curved spacetime manifold were discussed in \cite{mv}.
The relations that describe the duality actually give a local
parametrization of spacetime in terms of solutions of the Kaluza-Klein
equation and prepotentials.  If one would try to express the spacetime
measurements in terms of coordinate-field duality, one should also include
the gravitational phenomenon in this picture. Thus, it is natural to ask
what is the representation of the Einstein equations in this parametrization.
This is the aim of the present letter. Also, we will discuss the case when
the fields are quantized and we will analyse the local consequences of
quantum coordinate-field duality on the metric of spacetime.

In order to derive the Einstein equations from coordinate-field duality we
firstly note that there is a relationship between one-forms on spacetime
manifold $M$ and the differentials of the fields \cite{fm}. This relationship
is induced diffeomorphism like and it allows us to write the metric, the
metric connection, the curvature and the Ricci tensor in terms of fields and
prepotentials.\footnote{The map between one-forms on spacetime and
differentials of fields is not an induced diffeomorphism
since the duality involves prepotentials beside coordinates and fields. That
is a consequence of the fact that in local Cartezian coordinates there are
two linearly independent solutions of the Klein-Gordon equation on each
direction.} A second remark is that we can always pick up a local Cartezian
coordinate system, according to the principle of equivalence, and define
the coordinate-fields duality in this reference frame. Any other local
coordinate system can be obtained from the Cartezian one by a coordinate
transformation. The advantage of choosing local Cartezian coordinates is
that in this case we can use the simple duality relations given in \cite{fm}.
The price to be paid for this simplicity is that the construction is purely
local since, in general, a Cartezian coordinate system fails to exist globally.

Let us start with a spacetime manifold $M$ endowed with a metric fields $g$
and a scalar field $\phi$ that satisfies the Klein-Gordon equation. In an
open neighbourhood $U$ of a point $P \in M$ we pick up a local Cartezian
coordinate system $ \{ x^{\a} \}, \a = 0,1,\ldots ,n-1$ in which the metric is
diagonal $g_{\a \b }(x) = \eta_{\a \b}(x)$ and the Klein-Gordon equation takes
the form $ ( \Box_x + m^2) \phi(x)=0$. Then according to \cite{fm} the following
duality relations hold in $U$
\be
\frac{\sqrt{2m}}{\hbar} x^{\a} = \frac{1}{2}\p
\frac{\6 \F [\p ]}{\6\p } - \F
\label{dualrel}
\ee
for $ \a = 0,1,\ldots,n-1$. Here, $\F [\p ]$ are the prepotentials defined by
\be
\tp = \f{\6 \F [\p ]}{\6 \p}
\label{defprep}
\ee
and $\p$ and $\tp$ are two linearly independent solutions of the Klein-Gordon
equation that depend on the coordinates $x^{\a}$. The coordinates $x^{\b}$
with $\b \neq \a$ enter $\p$ and $\tp$ as parameters. The prepotential
$\F$ satisfies a nonlinear differential
equation in $U$  \cite{fm}
\be
4{\cal{F}}^{(\alpha )'''} + [V^{(\a )}(x^{\a}) + m^2 ](\p
{\cal{F}}^{(\alpha )''} - {\cal{F}}^{(\alpha )'} )^3 = 0 ,
\label{eqf}
\ee
where $ '= \6 /\6 \p$
and the potential $V^{(\a )}(x^{\a})$ has the following form
\be
V^{(\a )}(x^{\a}) = [\f{1}{\phi (x)}\sum^{n-1}_{\b = 0, b\neq \a}
\6^{\b}\6_{\b}\phi (x) ]|_{x^{\b\neq\a}fixed}.
\label{potential}
\ee
As a consequence of (\ref{dualrel}) the following relations between the
derivatives and differentials with respect to $\{ x^{\a} \}$ and
$\{ \p \}$, respectively, hold
\be
\f{\6}{\6 x^{\a}}  =  \f{(8m)^{\f{1}{2}}}{\hbar}\f{1}{\E}\f{\6}{\6 \p}
\label{derrel}
\ee
\be
d x^{\a}  =  \f{\hbar}{(8m)^{\f{1}{2}}} \E d\p ,
\label{difrel}
\ee
where $\E = \p {\cal{F}}^{(\alpha )''} - {\cal{F}}^{(\alpha )'}$. The
relations (\ref{derrel}) and (\ref{difrel}) represent an induced
parametrization on the spaces $T_{P}(U)$ and $T^{*}_{P}(U)$, respectively.
One should keep in mind for later manipulations that there is no summation
over $\a$ in the r.h.s. of (\ref{derrel}) and (\ref{difrel}). Using the
linearity of the metric tensor field (see for example \cite{dnf}) it is
easy to see that the components of metric in the $\{ (\p , \F ) \}$
parametrization are given by
\be
G_{\a \b} (\phi ) = \f{\hbar^{2}}{8m}\E E^{(\b )} \eta_{\a \b} (x).
\label{metphi}
\ee

Now let us take a general coordinate system
$ z^{\mu }$, $\mu = 0,1,\ldots ,n-1 $ in $U$, and let us denote the coordinate
transformation matrices by
\be
A^{\a}_{\mu } = \f{\6 x^{\a}}{\6 z^{\mu }} ~~,~~(A^{-1})^{\mu}_{\a} =
\f{\6 z^{\mu }}{\6 x^{\a }}.
\label{trmat}
\ee
The components of the metric in the new coordinate system are given by
\be
g_{\mu \nu}(z) = \f{8m}{\hbar^{2}}\f{1}{\E E^{(\b )}} A^{\a}_{\mu }A^{\b}_{\nu }
G_{\a \b}(\phi ),
\label{metz}
\ee
where the summation over $\a$ and $\b$ is performed. The components of metric
connection can be computed using the formula
\be
\Gamma^{\rho}_{\mu \nu} (z) = \f{1}{2}g^{\rho \sigma}(z)\sum_{P} \epsilon_{P}
P[\frac{\6 g_{\sigma \nu}(z)}{\6 z^{\mu }}],
\label{metcon}
\ee
where $P$ is a cyclic permutation of the ordered set of indices
$\{ \sigma \nu \mu \}$
and $\epsilon_{P}$ is the signature of $P$. By the coordinate transformation
(\ref{trmat}) the function $\p$ depends on all coordinates $z^{\mu}$. It is
easy to see that one can express the metric connection (\ref{metcon}) in the
$\{ (\p , \F ) \}$ parametrization and the result is given by the following
relation
\be
\Gamma^{\rho}_{\mu \nu} = (\f{2m}{\hbar})^{\f{1}{2}}
\f{E^{(\rho )}E^{(\sigma )}}{E^{(\gamma )}}(A^{-1})^{\rho}_{\tau}
(A^{-1})^{\sigma}_{\chi}
G^{\tau \chi}\sum_{P}\epsilon_P P[A^{\gamma}_{\mu}\f{\6}{\6 \phi^{(\gamma )}}
(\f{1}{\E E^{(\b )}}A^{\a}_{\sigma}A^{\b}_{\nu}G_{\a \b})].
\label{metcomp}
\ee
Using the definition of the curvature tensor field \cite{dnf}, it is
straightforward to compute the components of curvature in the
$\{ (\p , \F ) \}$ parametrization. For the sake of clarity, let us introduce
the following notation
\be
C^{(\gamma)}_{\mu \sigma \nu } = \sum_{P} \epsilon_{P}
P[A^{\gamma}_{\mu}\f{\6}{\6 \phi^{(\gamma )}}(\frac{1}{\E E^{(\beta )}}
A^{\a}_{\sigma}A^{\b}_{\nu}G_{\a \b})].
\label{ccc}
\ee
Then a little algebra gives the components of the curvature tensor
\bea
R^{\rho}_{\mu \nu \lambda} & = & \f{16m^2}{\hbar^2}\f{1}{E^{(\rho )}}
A^{\delta}_{\lambda}\f{\6}{\6\phi^{(\delta )}}
[\f{E^{(\rho )}E^{(\sigma )}}{E^{(\gamma )}}(A^{-1})^{\rho}_{\tau}
(A^{-1})^{\sigma}_{\chi}G^{\tau \chi}C^{(\gamma )}_{\mu \sigma \nu}]
\nonumber\\
&+&
\f{4m^2}{\hbar^2}\f{E^{(\rho )}E^{(\xi )}E^{(\sigma )}
E^{(\xi ')}}{E^{(\gamma )}E^{(\gamma ')}}
(A^{-1})^{\rho}_{\tau}(A^{-1})^{\xi}_{\chi}(A^{-1})^{\sigma}_{\tau '}
(A^{-1})^{\xi '}_{\chi '}G^{\tau \chi}G^{\tau '\chi '}
C^{(\gamma )}_{\sigma \xi \lambda}C^{(\gamma ')}_{\mu \xi '\nu}
\nonumber\\
&-&
(\lambda \leftrightarrow \nu ).
\label{curvten}
\eea
In the same way one can compute the components of the Ricci tensor and
the scalar curvature in terms of fields and prepotentials. To simplify the
results we introduce some notations as follows
\bea
\Sigma^{\rho \sigma}_{(\gamma )} &=&\f{E^{(\rho )}E^{(\sigma )}}{E^{(\gamma )}}
(A^{-1})^{\rho}_{\tau}(A^{-1})^{\sigma}_{\chi}G^{\tau \chi}
\nonumber\\
\Omega^{\rho \xi \sigma \xi '}_{(\gamma \gamma ')} &=&
\f{E^{(\rho )}E^{(\xi )}E^{(\sigma )}E^{(\xi ')}}{E^{(\gamma )}E^{(\gamma ')}}
(A^{-1})^{\rho}_{\tau}(A^{-1})^{\xi}_{\chi}(A^{-1})^{\sigma}_{\tau '}
(A^{-1})^{\xi '}_{\chi '}G^{\tau \chi}G^{\tau '\chi '}
\nonumber\\
\Lambda^{\mu '\lambda '}_{\mu \lambda } & = &
\f{\E E^{(\b )}}{E^{(\a ')}E^{(\b ')}}A^{\a '}_{\mu }A^{\b '}_{\lambda}
(A^{-1})^{\lambda '}_{\a}(A^{-1})^{\mu '}_{\b}G_{\a '\b '}G^{\a \b}
\nonumber\\
\Theta_{\mu \lambda} &=& \f{1}{E^{(\delta )}}A^{\delta}_{\lambda}
\f{\6}{\6\phi^{(\delta )}}
(\Sigma^{\rho \sigma}_{(\gamma )}C^{(\gamma )}_{\mu \sigma \rho})
+\f{1}{4}\Omega^{\rho \xi \sigma \xi '}_{(\gamma \gamma ')}
C^{(\gamma )}_{\sigma \xi \lambda}C^{(\gamma ')}_{\mu \xi '\rho}
-(\lambda '\leftrightarrow \rho )
\label{morenot}
\eea
With (\ref{morenot}) the local form of the Einstein equations with a vanishing
energy-momentum tensor in the $\{ (\p , \F ) \}$ parametrization is given by
\be
\Theta_{\mu \lambda} - \f{1}{2}\Lambda^{\mu '\lambda '}_{\mu \lambda }
\Theta_{\mu '\lambda '} = 0.
\label{einsteq}
\ee
Note that all the quantities in (\ref{einsteq}) depend on the scalar fields
$\p$ and on the prepotentials $\F$ which were defined in local Cartezian
coordinates. Moreover, using (\ref{eqf}) one can see that the potential
(\ref{potential}) determined by the other directions $x^{\b}$ also enter
(\ref{einsteq}) as expected. The equation (\ref{einsteq}) is given in the
general local coordinates $\{ z^{\a} \}$ in which the components of various
tensors were computed. One can obtain (\ref{einsteq}) in any other coordinate
system by replacing the matrices (\ref{trmat}) with the coresponding ones.
Using (\ref{derrel}) and (\ref{difrel}), one can express any spacetime tensor
in terms of fields and prepotentials locally as we did for the above tensors.
Thus, if the energy-momentum tensor does not vanish, one should add in the
r.h.s. of (\ref{einsteq}) the following term
\be
16\pi\f{mk}{\hbar^2}\f{1}{\E E^{(\b )}}A^{\a}_{\mu }A^{\b}_{\nu}{\cal{T}}_{\a \b}
\label{emten},
\ee
where $k$ is the Newton's constant in $n$ dimensions and ${\cal{T}}_{\a \b}$
are the components of the energy-momentum tensor in terms of $\p$.

If the Klein-Gordon field is quantized, the duality between coordinates and
fields induces a parametrization of flat spacetime in terms of field
operators \cite{fm,mv1}. In particular, we have an operatorial representation
of the coordinates and of the metric. In the presence of gravity, the problem
is in general more complicate. Nevertheless, one can show that the duality
between Cartezian coordinates and the quantum fields holds at least locally if
some constraints are imposed on the spacetime manifold \cite{mv}. Now let us
examine the local parametrization of the metric in terms of quantum fields in
the presence of gravity. To this end, we will take the linear independent
local solutions of the Klein-Gordon equation along each $x^{\a}$ of the
following form
\bea
{\phi}^{(\a )}_k(x^{\a }) = a_k \varphi^{(\a )}(k,x^{\a })
\nonumber\\
{\tilde{\phi}}^{(\a )}_k(x^{\a }) = a^{\dagger}_k
{\tilde{\varphi}}^{(\a )}(k,x^{\a }),
\label{fis}
\eea
where $a_k$ and $a^{\dagger}_k$ are annihilation and creation operators, respectively,
on the Fock space ${\cal{H}}_U$ defined locally on $U \in M$ \cite{bd}. The functions
$\varphi^{(\a )}(k,x)$ and ${\tilde{\varphi}}^{(\a )}(k,x)$ are linearly
independent solutions of the Klein-Gordon equation. As in the classical case,
they depend on the variable $x^{\a }$ while $x^{\b }$ for $\b \neq \a$
are treated as parameters. The index $k$ stands for all the indices necessary
to label an
independent mode of the field \cite{bd,mv}. For each local direction $ \a$
and each mode $k$ we introduce a prepotential ${\hat{\F}}_{k}
[\phi^{(\a )}]$ defined as in the classical case. In this setting, the duality
between spacetime coordinates and fields (\ref{fis}) is given by the
following relation
\bea
\f{\sqrt{2m}}{\hbar} {\hat{X}}^{\a }_k & = & \f{1}{2}\f{\6 {\hat{\F}}_{k}}
{\6 \phi^{(\a )}_k}\phi^{(\a )}_k -
{\hat{\F}}_{k} + {\hat{C}}^{(\a )}_{k}
\nonumber\\
{\hat{X}}^{\a }_k & = & a^{\dagger}_k a_k x^{\a}
\label{dualoper}
\eea
where the hat denotes operators. ${\hat{C}}^{(\a )}_{k}$ is an integration
constant and therefore it does not depend on $x^{\a}$ but only on the
parameters. The duality (\ref{dualoper}) induces corresponding
parametrizations of the tangent and cotangent spaces to $U$ in terms of the
variations of the quantum fields as in (\ref{derrel}) and (\ref{difrel}),
respectively. To compute these variations, we assume that the
operator ${\hat{E}}^{(\a )}_{k} = ({\hat{\F}}_k'' \phi^{(\a )}_k - 
({\hat{\F}}_k')/2$ is invertible. Then the quantum counterparts of
(\ref{derrel}) and (\ref{difrel}) have the following form
\bea
\f{\6 }{\6 {\hat{X}}^{\a }_k } & = &
\f{\sqrt{2m}}{\hbar}\f{\6 }{\6 {\phi}^{(\a )}_k} 
{\hat{E}}^{(\a )(-1)}_{k}
\nonumber\\
d {\hat{X}}^{\a }_k & = & \f{\hbar}{\sqrt{2m}}{\hat{E}}^{(\a )}_{k}
d {\phi}^{(\a )}_k ,
\label{paramquant}
\eea
where we consider that the differential and the derivative of the operators
${\hat{X}}^{\a }_k$ satisfy the following relation
\be
d {\hat{X}}^{\a }_k (\f{\6  }{\6 {\hat{X}}^{\a }_k })=
a^{\dagger}_k a_k dx^{\a}(\f{\6  }{\6 {\hat{X}}^{\a }_k })=1.
\label{difder}
\ee
Let us focus on a mode $k$ and drop the
corresponding index. We define the operator $d{\hat{S}}^2$ 
through the following relation
\be
d{\hat{S}}^2 = {\hat{N}}^2ds^2,
\label{opemetric}
\ee
where $\hat{N} = a^{\dagger}a$ and $ds^2$ is the local Minkowski metric. It is
easy to see that (\ref{opemetric}) can be cast into to following form
\be
d{\hat{S}}^2 = \f{\hbar^2}{4m}\eta_{\a \b}
\hat{E}^{(\a )}d{\hat{\phi}}^{(\a )}d^*({\tilde{\varphi}}^{(\b )}
\varphi^{(\b )})
+\f{\hbar^2}{4m}\eta_{\a \b}
\hat{E}^{(\a )}
\hat{E}^{(\b )}d{\hat{\phi}}^{(\a )}
d{\hat{\phi}}^{(\b )},
\label{opmetricext}
\ee
where we have used the following notation in the first term
\be
d^*({\tilde{\varphi}}^{(\b )}\varphi^{(\b )})
= d{\tilde{\varphi}}^{(\b )}{\varphi}^{(\b )}
-{\tilde{\varphi}}^{(\b )}d{\varphi}^{(\b )}.
\label{stard}
\ee
Since the functions ${\tilde{\varphi}}^{(\b )}$ and $\varphi^{(\b )}$ depend on
a single variable one can show that (\ref{stard}) is actually equal to
$(-\sqrt{8m}/\hbar) dx^{\b}$. Using this relation in (\ref{opmetricext}) one
can obtain the relation defining the metric in the operator parametrization
\be
\hat{N}(\hat{N} + 1){\hat{G}}_{\a \b} =
\f{\hbar^2}{2m}\eta_{\a \b}{\hat{E}}^{(\a )}{\hat{E}}^{(\b )}.
\label{mmetricop}
\ee
The expression above corresponds to the local flat metric obtained from a
local Cartezian system. If we want to pass to a general coordinate system,
we have to perform a coordinate transformation (\ref{trmat}). In general 
nor the vacuum of the theory neither the creation and annihilation
operators are invariant under such of transformation and thus thermal particles
can be created \cite{bd}. However, if we consider coordinate transformation
that generate a small deformation of the metric so that the thermal effects
can be disregarded
on the local Fock space and on operators, we see that the deformed metric in
the
coordinate system $\{ z^{\mu } \}$ is given by the following relation
\be
{\hat{G}}_{\mu \nu } = \f{2m}{\hbar^2}A^{\a}_{\mu }A^{\b}_{\nu}
({\hat{E}}^{(\a )} {\hat{E}}^{(\b )})^{-1}\hat{N}(\hat{N}+1)\hat{G}_{\a \b}
\label{finmet}.
\ee

The relations obtained above are true for any mode $k$ and for any direction
$\a$ in the open $U$. Since the modes are independent from each other we
see that the various representations of the coordinates given by
(\ref{dualoper}) commute with each other. Therefore, the geometry that can
be constructed from these representations should be commutative. The fact
that it is possible to parametrize locally the spacetime in terms of field
operators gave us an operatorial representation of the metric in
(\ref{mmetricop}) and (\ref{finmet}). If one picks up an orthogonal
basis formed by
the eigenvectors of $\hat{N}_k$ denoted by $\{ |k,n \rangle \}$
by using (\ref{opmetricext}) one
can show that the following relation holds
\be
ds^2 = \f{\hbar^2}{16m}
\sqrt{n(n-1)}
\langle k,n|
({\hat{\F}}_k'' \phi^{(\a )}_k - 
({\hat{\F}}_k')({\hat{\cal{F}^{(\b )}}}_k'' \phi^{(\b )}_k - 
({\hat{\cal{F}^{(\b )}}}_k')aa|k,n-2 \rangle
\eta_{\a \b}d\varphi^{(\a )}d\varphi^{(\b )}
\label{quantfluc}.
\ee
The relation (\ref{quantfluc}) depends on the field operators defined in
$P$ as well as on the prepotentials. One can obtain a time-like, space-like
or light-like line element in the r.h.s. of (\ref{quantfluc}) depending on
the action of these operators on the states of $\hat{N}$ in the neighbourhood
of $P$. In general, the prepotentials that enter the duality between
coordinates and fields should satisfy a differential equation in both classical
and quantum case \cite{mv}. However, from (\ref{quantfluc}) we can see that
there are further constraints on the prepotentials as a consequence of the
fact that the metric can be expressed in terms of quantum field operators.

To summarize, we have obtained the local expressions of the Einstein
equations in a local parametrization of space-time that involves
local solutions of the Klein-Gordon equation and the prepotentials.
The main result is given by (\ref{einsteq}) which is the dual form of the
Einstein equation in the sense of \cite{fm}. In order to obtain this equation we have
used a local Cartezian coordinate system. Also, we have investigated the
local structure of the spacetime metric in this parametrization when the
fields are quantized. Locally, there is an operatorial description of the
Cartezian coordinates of spacetime and of the metric. In this description,
the character of the line element is given by the action of the field
operators and of the prepotential functionals on the field operators in the
neighbourhood of a point. This result is given by (\ref{quantfluc}).

The treatement of gravity within the framework of the duality proposed in
\cite{fm} as was given in this paper, is completely local. It would be
interesting to see if there is possible to formulate a global version of
this duality. Also, in this formulation the representation of the metric in
terms of quantum fields is a consequence of the quantum counterpart of the
coordinate-field duality. It would be interesting to see if there is a
dynamics of the metric in this context, similar to classical case. 
To this end one should try to understand better the quantum duality. The fact
that the cordinates can be represented in terms of various solutions of
field equations raises the natural question about the relations between
these representations and about the possibility of parametrizing the
spacetime in terms of various fields. For the discussion of some of these
problems we refer the reader to \cite{mv1,cgv}.

\acknowledgements

I would like to thank to M. C. B. Abdalla, M. A. De Andrade, A. L. Gadelha,
J. A. Helay\"{e}l-Neto for
useful discussions and to DPC-CBPF and LNCC for hospitality during the
preparation of the paper and to the referee for suggesting me to study the
second part of this work. I also acknowledge a FAPESP postdoc fellowship.


\begin{references}

\bibitem{fm} A. E. Faraggi and M. Matone, Phys. Rev. Lett. {\bf 78}, 163
(1997).

\bibitem{sw} N. Seiberg and E. Witten, Nucl. Phys. B {\bf 426}, 19 (1994).

\bibitem{bm} G. Bonelli and M. Matone, Phys. Rev. Lett. {\bf 76}, 4107 (1996).

\bibitem{mm} M. Matone, Phys. Lett. B {\bf 357}, 342 (1995).

\bibitem{af} A. E. Faraggi, hep-th/9910042. 

\bibitem{bfm} G. Bertoldi, A. E. Faraggi and M. Matone, hep-th/9909201.

\bibitem{lr} G. Landi and C. Rovelli, Phys. Rev. Lett. {\bf 78}, 3051 (1997);
Mod. Phys. Lett. A {\bf 13}, 479 (1998).

\bibitem{gl} G. Landi, gr-qc/9906044.

\bibitem{viv} I. V. Vancea, Phys. Rev. Lett. {\bf 79}, 3121 (1997),
Err. ibid. {\bf 80}, 1355 (1998); Phys. Rev. D {\bf 58}, 045005 (1998).

\bibitem{cv} C. Ciuhu and I. V. Vancea, gr-qc/9807011 to be
published in Mod. Phys. Lett. A.

\bibitem{pv} N. Pauna and I. V. Vancea, Mod. Phys. Lett. A {\bf 13}, 3091
(1998).

\bibitem{mv} M. A. De Andrade and I. V. Vancea,
Phys. Lett. B. {\bf 474}(2000)46

\bibitem{dnf} B. Doubrovine, S. Novikov and A. Fomenko,
{\it G\'{e}om\'{e}trie Contemporaine} (\'{E}ditions MIR, Moscou, 1979) 

\bibitem{mv1} M. A. De Andrade and I. V. Vancea, {\it Coordinate-Field Duality
and Quantum Field Theory}, in preparation

\bibitem{bd} N. D. Birrell and P. C. Davies, {\it Quantum Fields in Curved
Space} (Cambridge, 1989)

\bibitem{cgv} M. C. B. Abdalla, A. L. Gadelha and I. V. Vancea, hep-th/0002217 

\end{references}
\end{document}